\documentclass[twocolumn,amssymb,amsmath,nofootinbib,tightenlines,showpacs,floatfix,superscriptaddress,aps,prb]{revtex4-1}
\usepackage{amsfonts}
\usepackage{bm}%
\usepackage[colorlinks=true,linkcolor=blue,filecolor=blue,menucolor=yellow,urlcolor=blue,citecolor=blue,anchorcolor=blue]{hyperref}
\usepackage{graphicx}
\usepackage{dcolumn}
\usepackage{rotating}

\begin{document}

\title{Ising Model on a random network with annealed or quenched disorder} 

\author{Abdul N. Malmi-Kakkada}
\email{malmikakkada@physics.umn.edu}
\author{Oriol T. Valls}
\email{otvalls@umn.edu}
\altaffiliation{Also at Minnesota Supercomputer Institute, University of Minnesota,
Minneapolis, Minnesota 55455}
\affiliation{School of Physics and Astronomy, University of Minnesota, 
Minneapolis, Minnesota 55455}

\author{Chandan Dasgupta }
\email{cdgupta@iisc.ernet.in}
\altaffiliation{Also at Condensed Matter Theory Unit, Jawaharlal Nehru Centre
for Advanced Scientific Research, Bangalore 560064, India}
\affiliation{Centre for Condensed Matter Theory, Department of Physics, 
Indian Institute  of Science, Bangalore 560012, India}

\date{\today}

\begin{abstract} 
We study the equilibrium properties of an Ising model
on a disordered random network where 
the disorder can be  quenched or annealed. 
The network consists of  four-fold
coordinated sites connected via variable length
one-dimensional chains. Our emphasis is on
nonuniversal properties and
we consider 
the transition temperature and 
other equilibrium thermodynamic
properties, including those associated with
one dimensional fluctuations arising from the chains. 
We use analytic methods in the annealed case,
and a Monte Carlo simulation  for 
the quenched disorder. Our objective is to
study the difference between quenched and annealed results  
with a broad random distribution of interaction parameters.
The former represents a situation where the time scale associated
with the randomness is very long and the corresponding
degrees of freedom
can be viewed as frozen, while the
annealed case models the situation where this is not so.
We find that the transition temperature and the entropy associated
with one dimensional fluctuations are always higher for 
quenched disorder than in the annealed case. 
These differences increase 
with the strength of the disorder
up to a saturating value. 
We discuss our results in connection to  physical
systems where a broad distribution of interaction strengths 
is present.
\end{abstract}


\maketitle

\section{Introduction} 
\label{Intro}
Spin models on random networks are relevant to many physical
phenomena and therefore have been studied in a variety of contexts. 
Early studies of phase transitions in spin models on random networks 
were concerned with the critical behavior of randomly diluted magnetic
systems.~\cite{review1,review2} The system-spanning percolation  
cluster~\cite{percolation} just above the
percolation threshold has a ramified network structure with fractal dimension
less than the physical dimension of the system: hence it is necessary to work out 
the critical behavior of spin models defined on a random network to 
develop an understanding of phase transitions in dilute magnets near 
the percolation point. The well-known ``node-link-blob''
descriptions of percolation clusters~\cite{nodes-link1,nodes-link2} were 
developed to address this problem. 
Spin models on artificially constructed regular fractal networks were
also studied:~\cite{fractal1,fractal2} an advantage of these models is
that their equilibrium  
thermodynamic properties could 
be calculated exactly for some of the relevant networks. Also, such studies were
expected to provide some insight on the behavior of spin systems on real
fractal networks.

More recently, there has been an explosion of research activity on random
networks that are believed to describe various systems of interest in physics, 
biology, engineering
and social 
sciences.~\cite{network1,network2,network3,network4,network5} 
Some of these studies have 
concentrated on structural aspects of random 
networks,~\cite{small-world,scale-free,recovery} while others 
have investigated the collective behavior of interacting objects residing on 
different kinds of random
networks of interest. Models in which spin variables defined on random networks 
interact with
one another provide examples of 
systems that exhibit nontrivial 
collective behavior, such as phase transitions.~\cite{phase-transition} For this reason,
a variety of models with Ising,~\cite{ising1,ising2,ising3,ising4,ising5} 
Potts~\cite{potts1,potts2} and\cite{xy,xy1} XY 
spins, defined on different kinds of random networks, have been studied 
in recent years using
both analytic and numerical methods. These studies have revealed many 
interesting features\cite{lima} in the 
equilibrium and dynamic behavior of spin systems on random networks.

Disorder is an essential aspect of spin models defined on random networks. 
Depending on the 
network being considered, disorder may appear in different aspects of 
the spin model, such as  
in the number
of spins interacting with a particular one 
(the degree of connectivity may be 
different~\cite{appo1,appo2} for different nodes at 
which the spins are located) and the strength of the interaction between pairs
of spins (the interaction strength may be different for spin pairs in the 
network that are separated by
different distances). The disorder in such systems, arising from the randomness 
in the structure of the
network, is generally assumed to be {\it quenched} in the sense that
for any realization of the model the thermodynamic degrees of freedom 
associated with 
the network structure are fixed, and therefore 
the network does not evolve in time. In theoretical treatments of the 
equilibrium behavior of such
spin systems, the free energy is therefore averaged over different realizations 
of the disorder.~\cite{review1} 
However, the validity of the assumption of the disorder being 
quenched depends crucially on the comparison of
relative time scales - real networks do evolve in time and the
assumption of quenched disorder would not be valid unless the time scale over 
which the network changes is orders of magnitude larger than 
the time scale of the spin
fluctuations. If these two time scales are comparable to each other, or at least
not too different, 
then the disorder should be considered
to be {\it annealed} and the partition function of the spin system (not the free energy) 
should be thermodynamically averaged 
over different realizations of the disorder in the network, to obtain a 
correct theoretical description of the
equilibrium behavior. Thus, the disorder in the spin system would change from 
quenched to annealed if the 
time scale for the evolution of the network structure decreases 
from being much longer than that for spin fluctuations 
to values roughly comparable to, or shorter, than the typical relaxation time of the 
spin variables. 
In this paper, we address, 
within the context of a simple specific
model, the question of how the equilibrium behavior of a
disordered spin system (specifically, its behavior near a phase transition) 
would be affected by such a change in the dynamics of the network on which 
the spins reside, so that 
the fluctuations
associated with the disorder would have to be properly
included in the thermodynamics calculations.

The question of how disorder affects the critical behavior near a phase 
transition has been extensively  
studied. Here, we consider 
spin systems in which the disorder does not introduce frustration
as it might  arise, for example, 
from the presence of both ferromagnetic and antiferromagnetic interactions. 
In  such systems, the presence of quenched  
disorder changes the universality class of the phase transition if 
the specific heat exponent for the transition in the
system without disorder is positive (the Harris criterion~\cite{harris}). 
The presence of annealed disorder usually does not
change the universality class of the phase transition because one recovers an 
effective model without disorder after
averaging the expression for the partition function over the disorder variables 
(in some cases, the presence of
annealed disorder leads to a ``Fisher renormalization''~\cite{fisher-renorm} of 
the critical exponents). 

A question that has 
not received much 
attention  in the  
recent literature, although touched upon in 
some older work,\cite{thorpe,falk,thorpe1} is how the transition temperature 
itself, and other  thermodynamic quantities, 
are affected as the nature of the
disorder is changed from quenched to annealed, reflecting a difference in 
the network dynamics. This is one of the 
main issues addressed in the present study. The answer to this question is not 
universal - it depends on the 
specifics of the system being considered. 
Earlier studies~\cite{thorpe,thorpe1} considered disordered spin
models in which the distribution of the interaction parameter is narrow,
such as magnetic systems with bond dilution in which the interaction
parameter can have two values,  $J$ and $0$, 
and models in 
which it has a Gaussian distribution
with width much smaller than the average. 
These studies show that the thermodynamic behavior and the 
transition temperatures
of quenched and annealed systems 
are  similar. In contrast, in the model we consider here 
the distribution of the interaction parameter is very broad 
(log-normal, see below). 
In such cases the differences between quenched and 
annealed properties  with this kind of disorder 
have not been  
previously analyzed in any detail. 
We give below examples of 
systems for which  
this question is relevant -- our work was partly motivated by these problems,
although it is quite independent of them. 

The possibility of supersolid behavior~\cite{ssrev} in  $^4$He 
arising from superfluidity along a random network
of dislocation lines has been considered~\cite{mc1,mc2,shevchenko,dgt,toner,us1} 
recently. 
Quantum Monte Carlo
simulations~\cite{mc1,mc2} have shown that superfluidity can occur near the 
core of a 
dislocation line in solid $^4$He or 
along grain boundaries.\cite{us1} 
The transition in a model in which superfluidity 
occurs near dislocation lines has been
investigated~\cite{shevchenko,dgt,toner} theoretically, assuming a 
frozen dislocation network 
(quenched 
disorder). However, dislocation line segments do fluctuate in time 
and it 
has been suggested~\cite{balibar}
that this motion 
may 
suppress the local temperature of 
superfluid ordering. 
Since the  dislocation motion changes the 
nature of the disorder in the superfluid problem (described by a ferromagnetic
XY model) from quenched to annealed, a relevant question 
is how the nature of the disorder affects the transition
temperature. Although the initial experiments~\cite{chan1} 
on supersolidity are 
now believed~\cite{chan2} to reflect an 
elastic anomaly, 
the question of how the
motion of dislocation line segments affects superfluid ordering 
is important 
because of the occurrence of 
supersolid behavior
arising from superfluidity along a network of defects has been 
established in 
numerical studies.~\cite{mc1,mc2,us1} 
The effective ferromagnetic interaction between 
superfluid variables located at nearest-neighbor nodes of a 
disordered dislocation network 
falls off exponentially~\cite{shevchenko} 
with 
the length of the network segment that connects the nodes. 
If the nodes are distributed randomly in space, then  this
effective interaction would be a random 
variable with a very broad distribution. 

More generally, there are 
other systems of interest~\cite{sarma1,sarma2,boris,ravin} where the effective
interaction between neighboring spins is a random variable 
with a broad distribution. A system of this
kind 
that has received a lot of attention in recent years 
is dilute magnetic semiconductors~\cite{sarma1,sarma2} 
in which 
spins of localized holes interact ferromagnetically via the
spins of magnetic impurities present in the system. 
The quenched disorder here 
arises from the random locations of the holes, with the interaction
strength falling off exponentially with the distance between two holes. 
This results in a broad distribution of interaction strength -- an essential
feature of the  model we study here. 
There is no reliable 
analytic method for calculating the transition
temperature and thermodynamic properties of such quenched systems. 
A comparison of the properties of quenched and annealed 
versions of such models would be very useful: 
analytic calculations of the properties with annealed
disorder 
are possible 
because  
they can be mapped exactly~\cite{thorpe} to models without disorder.
If 
the properties of quenched and annealed versions of
models with a very broad distribution of the interaction strength were 
similar, 
(as in~\cite{thorpe,thorpe1} the case of a narrow distribution of
the interaction strength), 
then an analytic calculation of 
the properties of the annealed model 
would be broadly valid for the physically relevant quenched model.  
The spin model we study here provides a simple example of disordered 
systems with a broad distribution of the 
interaction strength. 

In this paper, we have studied the thermodynamics of 
a disordered ferromagnetic spin model defined on a two-dimensional 
random network,
with emphasis on how 
the thermodynamics, including the transition temperature, 
is affected by a change in the nature 
(quenched or annealed) of the disorder. 
For simplicity, we consider  Ising spins 
(instead of XY spins which would be appropriate for describing
superfluid ordering). 
The network is assumed 
to have the same connectivity as the square lattice, 
i.e. every node is connected to four other nodes. 
Ising spins are defined both at these four-fold coordinated 
nodes and on the links that connect them. 
Spins on these one-dimensional links are placed uniformly 
so that the number of spins on a link is equal to its
length measured in units of the spacing between nearest-neighbor sites. 
Each Ising spin (whatever its coordination number) 
interacts  ferromagnetically with its nearest 
neighbors. The disorder arises from a distribution 
of the lengths of the one-dimensional links, i.e. the 
number of spins in these links. 
In the dislocation network 
problem, this distribution may arise from 
roughening of dislocation line segments.~\cite{roughening} 
We assume a Gaussian distribution for the number of spins 
in each link, 
and  study the thermodynamic 
behavior for different values of the average and 
standard deviation of this distribution. 
Since the effective interaction between two spins  at 
nearest-neighbor nodes falls off exponentially
with the number of spins in the link that joins these nodes (see below), a 
Gaussian distribution of the number of spins
in a link implies a very broad, log-normal distribution for the effective 
interaction.
The thermodynamic behavior is studied
analytically for annealed randomness and via Monte Carlo 
simulations for quenched randomness. We find that the transition 
temperature with quenched disorder is always higher 
than that in the case of annealed disorder with the same distribution. 
This difference initially increases with the strength of the disorder,
and eventually saturates for 
larger values of a parameter that characterizes the disorder. 
For both cases, the specific 
heat as a function of temperature exhibits two peaks: 
a sharp one at the  phase transition and a rounded peak at a 
higher temperature, reflecting the
one-dimensional fluctuations along the links. 
The qualitative behavior of the specific heat (and the associated entropy) 
in both cases 
is very similar, but there are quantitative 
differences that become more pronounced as the strength of the disorder 
increases.

The rest of the paper is organized as follows. In section~\ref{methods}, we describe
in more detail the 
model under study and describe the methods we follow both
for  analytic calculations and simulations.
The results obtained from the study 
of this model and its relevance to the problems mentioned 
above are described in detail in
section~\ref{result}. Section~\ref{conclusion} 
contains a summary of the main results and concluding remarks.

\section{Model and methods} 
\label{methods}

\begin{figure}
\includegraphics[width=0.45\textwidth] {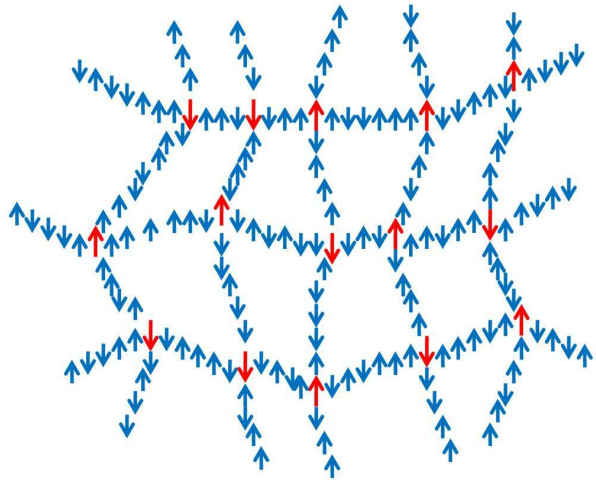} 
\caption{(Color online) Sketch of part of the
coupled Ising system under study. The (red) arrows at the
nodes are four-fold coordinated Ising spins. They are
connected by chains of Ising spins (blue). The chains have 
variable lengths.}
\label{fig0}
\end{figure}

To study the situations described 
in the Introduction, we consider a system of
two coupled Ising models.  It consists of a system
of four-fold coordinated Ising spins (a two-dimensional
system) connected by one-dimensional
chains of two-fold coordinated Ising spins. 
In the chains, each spin interacts with 
its two nearest neighbors, while
spins at the nodal sites (crossing points of the chains) interact 
with their four nearest neighbors. The scheme is  
illustrated in Fig.~\ref{fig0}.  
In this figure, the nodal spins, indicated by red color, interact with 
their four nearest 
neighbors, which
belong to four different chains, while the spins along the one dimensional 
chains, indicated
by blue color, interact with their two nearest neighbors.
In this simple network model, a distribution
in the number of 1D spins in the chains leads to 
randomness in the effective interaction between 2D spins. 

We will denote the two dimensional spins as $S_i$ where $i$ is a two
dimensional index running from 1 to $N^2$ where $N$ is a very large
number. 
The number, $n_{ij}$,  of spins in the  chain connecting sites $i$
and $j$ dictates the  effective 
``distance" between nodal spins. Selecting the set $n_{ij}$  randomly
according to some probability distribution (see below) leads to 
the realization of a random network of coupled spins. 
The model Hamiltonian can then be written as:
\begin{equation}
\label{modelh}
H = -J \sum_{i=1}^{N^2}\sum_{\alpha=1}^{4} S_{i} \sigma_{\alpha} - 
J \sum_{<ij>}\sum_{\alpha=1}^{n_{ij}-1} \sigma_{\alpha}\sigma_{\alpha+1} 
\end{equation} 
where $S_{i}=\pm 1$ and $\sigma_{\alpha}=\pm 1$ are 
the 2D and 1D spins respectively. 
The $\sigma_\alpha$ in the first term on the right
are those connected directly to $S_i$, 
and the first summation in the last term denotes
the sum over all chains, connecting 
neighboring sites $i$ and $j$. The quantity $J$ is
the exchange energy, which  we will set to unity in
most of the calculations
below. The above formula
assumes all $n_{ij}>1$. When one of the $n_{ij}=0$ the corresponding
$S_i$ and $S_j$ are connected directly. If all 
$n_{ij}=0$  we recover the
standard 2D Ising model result with a 
transition temperature of ${T_{c}}/{J}=2.26$ 
(we set $k_{B}=1$ throughout the paper). When one of the $n_{ij}=1$
Eq.~(\ref{modelh}) must be modified so that the term corresponding
to the chain connecting $S_i$ 
and $S_j$ is omitted.

The limit in which all chains are of
equal length, $n_{ij}\equiv n$, can easily be considered 
analytically. To do so, we first calculate the
partition function for a finite 1D chain.  
Starting with the Hamiltonian 
\begin{equation}
H_{1D} = -J \sum_{\alpha=1}^{n-1} \sigma_{\alpha}\sigma_{\alpha+1}, 
\end{equation}
it is easily shown from elementary transfer matrix methods 
that the entire system can be mapped onto an ordinary 
square lattice Ising model, with an
effective interaction, $J^{(n)}$, between two four-fold 
coordinated spins, separated by a 
``distance" of $n$ spins, given by:
\begin{equation}
\label{jeff}
\tanh(\frac{J^{(n)}}{T})=\tanh^{n+1}(\frac{J}{T}).  
\end{equation}
The free energy for the coupled Ising model at fixed $n$
can then be calculated based on the standard Onsager result, plus an 
additional contribution from the chains of 1D spins linking the 
nodal 2D spins. Setting henceforth $J=1$, 
the contribution to 
the free energy from the chains can easily be
shown to be (for one chain): 
\begin{equation} 
\begin{split}
{F_{1D}} = &-Tn \log(2) -T(n+1) \log(\cosh(\frac{1}{T})) \\
&+ {T}\log(\cosh(\frac{J^{(n)}}{T})) 
\label{f1}
\end{split}
\end{equation}
and the contribution from the 2D spins is:
\begin{equation} 
\begin{split} 
{F_{2D}} = &-{T}\log(2\cosh(\frac{2J^{(n)}}{T}))  \\ 
&-\frac{T}{2\pi}\int\limits_0^\pi \log(\frac{1}{2}(1+\sqrt{1-P^{2}\sin^{2}\phi}))\mathrm{d}\phi
\label{f2}
\end{split}
\end{equation}
where $P$ is defined as:
\begin{equation}
P \equiv \frac{2\sinh(\frac{2J^{(n)}}{T})}{\cosh^{2}(\frac{2J^{(n)}}{T})}.
\end{equation}
All thermodynamic quantities can then be calculated from the 
free energy. 
We will be interested in the behavior of
thermodynamic quantities such as the specific heat 
and the entropy ($S$)
since they are important  
in understanding how the behavior of the system 
near a phase transition is affected by changes in the dynamics 
of the network.
Other thermodynamic quantities such as the spontaneous magnetization and
the magnetic susceptibility can also be studied, but 
we will focus in this work on the entropy and its
derivatives.

In general, we are interested in the case where
the $n_{ij}$ vary from chain to chain. Accordingly, we generate
a random Ising network by choosing 
$n_{ij}$ for each chain from a
gaussian probability distribution:
\begin{equation}
\label{gaussian}
P(n_{ij})= \frac{e^{\frac{-(n_{ij}-\widetilde{n})^{2}}{2\delta^{2}}}}{\sqrt(2\pi)\delta}
\end{equation}
where $\widetilde{n}$ is the average of $n_{ij}$ (average number of 1D spins in a chain) 
and $\delta$ the standard deviation of the gaussian distribution.

Using this probability distribution, we will investigate, as explained
in the Introduction, how the thermodynamic behavior is affected by 
quenched and annealed disorder in the network. 
In the quenched case, the value of $n_{ij}$ in each individual
1D chain is fixed but it varies from 
one chain to the next according to the gaussian random distribution.
This serves as a proxy for 
a disordered network in which the characteristic time scale for 
changes in the network is much longer than the characteristic time scale
for spin fluctuations. 
For the annealed case, the values of $n_{ij}$ are allowed to thermally
fluctuate and this scenario serves as a proxy for a dynamic
network whereby the two characteristic time scales mentioned above 
are comparable to each other. 
In studying the differences
between quenched and annealed disorder,
we will focus on
features of the heat capacity
such as how the temperatures at which $C_v$ has peaks
(corresponding to 2D and 1D behavior, see below) 
change between the two scenarios.
Changes in the peak temperatures  
depending on the type of disorder will allow
us to address the question of the role
that the dynamics of the network plays 
in the ordering of the spins. 


When one treats  the disorder as annealed, 
the free energy 
of our system is:
 \begin{equation}
 \label{fanneal}
 F_{a} = -T\log\langle Z \rangle,
 \end{equation}
where the angular brackets denote an average over the 
gaussian probability distribution, Eq.~(\ref{gaussian}).
Therefore, $\langle Z({n_{ij}})\rangle$ needs to be calculated. 
For a gaussian distribution, this calculation can be done analytically.
By tracing over the 1D spins in the chains, the model
becomes one in which the 2D spins
occupying the nodes interact according to $J^{(n)}$ given in Eq.~(\ref{jeff}).  
Evaluating then  the average of the partition function over the 
gaussian distribution, the annealed Ising model is mapped onto an effective
ferromagnetic square lattice Ising model with equal interactions 
$J^{(\widetilde{n},\delta)}_{a}$, 
given by:
\begin{equation}
 \label{zanneal}
 \tanh(\frac{J^{(\widetilde{n},\delta)}_{a}}{T})  = 
 \frac{\langle \sinh(\frac{J^{(n)}}{T}) \rangle}{\langle \cosh(\frac{J^{(n)}}{T})
 \rangle},
   \end{equation}
where the average over the discrete gaussian probability distribution
indicated by the angular brackets can be easily performed.
Thus, the effective interaction for the annealed
model is simply  a function of $\widetilde{n}$ and $\delta$.
The annealed free energy can then be calculated 
based on a procedure similar to the case where $n$ is fixed 
in each link (no disorder) as presented in
Eqs.~(\ref{f1}-\ref{f2}). 

For a  system with quenched disorder, the randomness is frozen in each 
realization
of the network. 
We generate realizations of the network whereby 
the couplings $J^{(n_{ij})}$, satisfying Eq.~(\ref{jeff}), 
vary from node to node
according to the probability distribution in Eq.~(\ref{gaussian}) for $n_{ij}$.
This corresponds to a model on a regular lattice 
with a random distribution of couplings $J^{(n_{ij})}$.
The free energy in the quenched case takes the form:
 \begin{equation}
 \label{fquench}
 F_{q} =  -T\langle \log Z \rangle, 
 \end{equation}
where the angular brackets still represent an average over the distribution
of chain lengths. 
Since such a calculation is 
analytically intractable, we use
Monte Carlo (MC) simulations to study the thermodynamic behavior of
the model with quenched disorder. 
A standard MC procedure with Metropolis algorithm
is used in our study. 
In each run 
in the simulation, the heat capacity of the 
spin system can be obtained either from the 
fluctuations of the internal energy 
or by 
taking the derivative of $\bar{E}$, (the overbar denote MC averaging) 
the average energy
per spin, with respect to the temperature. 
By subsequently 
averaging the heat capacity over a sufficiently
large number of  realizations of the chain length configurations 
(over twelve realizations in this study)  
of $n_{ij}$, we obtain the heat capacity for the random
Ising model with frozen disorder.
Each of the 12 realizations is  characterized  
by a unique random set of the 
effective interaction ($J^{(n)}$) between 2D spins. 


For the study of any random network, 
it is important to be able to tune the level of disorder. 
For the current model, the randomness of the network 
can be controlled by adjusting the values of $\widetilde{n}$ and
$\delta$, with 
the limit $\delta \rightarrow 0$ recovering the fixed $n$ coupled
Ising model discussed above. Since the number of 1D spins in the
chains cannot be negative, we use values of $\delta$ and 
$\widetilde{n}$ such that ${\delta}/{\widetilde{n}} \le 0.5$. 
With this choice,  there is only a very small 
probability of obtaining negative values for $n_{ij}$. In such 
rare cases, in 
the MC simulation, we set
the number of 1D spins in those chains to be two. 

The quantity $\delta/\widetilde{n}$ can be used as a measure of the amount
of disorder present in the network. A convenient and 
more physical alternative way to characterize the disorder in this random
coupled field model, is via the standard deviation
of $J^{(n)}$. Thus, we define a parameter
\begin{equation}
\label{parameter}
k_{\delta} = \frac{\sqrt{\langle [J^{(n)}]^{2}\rangle - \langle J^{(n)}\rangle^{2}}}{\langle J^{(n)}\rangle}.
\end{equation}
The quantity $k_\delta$  (which depends
also on $\widetilde{n}$) quantifies the level of disorder
in the random Ising model 
in terms of the spread in the effective 
interaction between 2D spins. 
The  differences between the properties of 
quenched and annealed networks 
can be analyzed in terms of  either ${\delta}/{\widetilde{n}}$ or  $k_{\delta}$
with a wide range of values 
considered for both 
$\widetilde{n}$ and $\delta$.

\section{Results}
\label{result}
In this section we present the results of our study on 
the random Ising model. We start by briefly 
discussing  
the fixed $n$ model ($\delta = 0$, i.e. no disorder) and 
then proceed to the 
random  model with $\delta \neq 0$.  
For the random model, 
we analyze differences between 
quenched and annealed disorder 
for both components of the coupled field - 1D and 
2D - by studying the behavior 
of the
heat capacity and entropy. 
For the numerical (quenched) results,
we have simulated samples with the number of 2D spins ($ N \times N$) from
$16\times16$ to $30\times30$, with periodic boundary conditions. 
Even though the values of $N$ used in our simulation
are relatively small, the total number of spins, including
the 1D ones is much larger: 
for e.g. a sample network with $N = 20$ and 
$\widetilde{n} = 19$ the number of spins
is approximately
$N \times N + 2 \times N \times N \times \widetilde{n}
= 15600.$
Finite size effects
in the numerical simulation  
are also analyzed in order
to estimate the error margin associated with our 
results. These are indicated by error bars where warranted.

\subsection{No Disorder}
\begin{figure}
\includegraphics[width=0.5\textwidth] {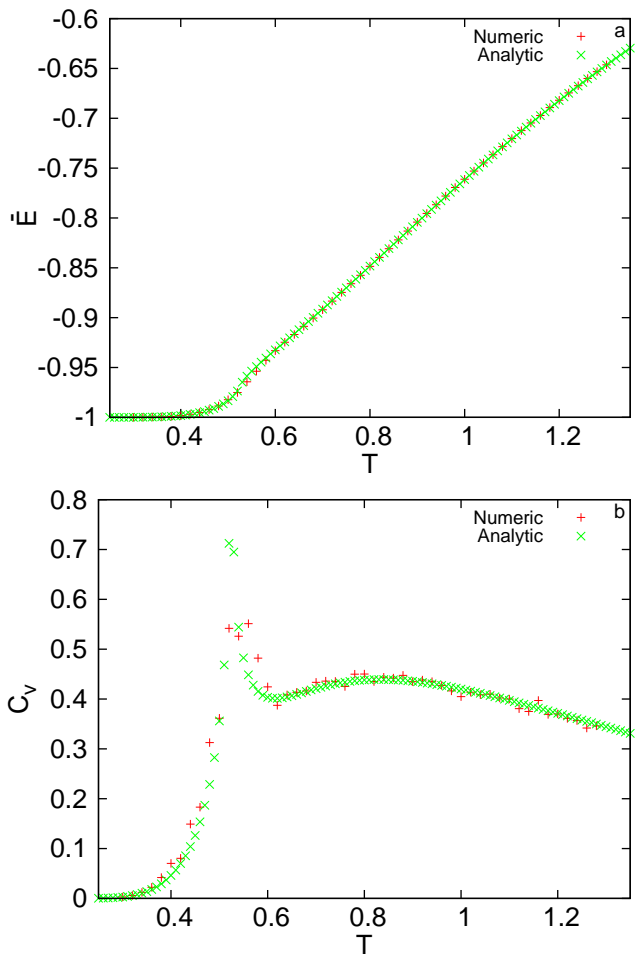}
\caption{(Color online) Comparison of analytic and MC
results. (a) Plot of the 
average energy per spin $\bar{E}$ vs temperature, 
for fixed chain length. The numerical 
results are for $N=16$ and $n=19$. (b) Plot 
of the corresponding heat capacity per spin
vs temperature. The parameters are the same as in part (a).}
\label{figener}
\end{figure}

For  fixed $n$,
the energy and heat capacity  can be
calculated analytically starting from the free energy expressions 
in the previous Section, Eqs.~(\ref{f1})-(\ref{f2}). 
Typical results for the temperature dependence 
of the energy and heat capacity
per spin are plotted in Fig.~\ref{figener}.
Since, when considering the  disordered ($\delta>0$) case below we will
have to take recourse, in part, to numerical methods, we have also
computed the same quantities numerically, to test the same 
numerical procedures
that will be employed later. These results are also plotted in 
Fig.~\ref{figener}. As mentioned above, the units of temperature
throughout this discussion are such that $J=1$.
The numerical results shown there are based on obtaining 
the average  $\bar{E}$ 
over a sufficiently large number of MC steps per spin: typically
about 16,000 turn out to be needed, 
and then 
numerically taking the derivative of this average with 
respect to temperature to obtain the heat capacity.
Despite the modest size of $N$ chosen for this display, it is clear
that the numerical results agree sufficiently well with the analytic results, 
thereby validating our numerical procedures.

\begin{figure}
\begin{turn}{-90}
\includegraphics[width=0.35\textwidth] {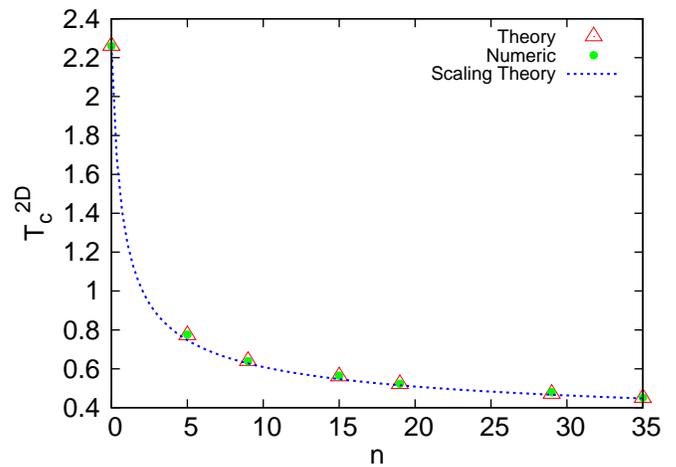}
\end{turn}
\caption{Plot of $T_{c}^{2D}$  vs. number of spins ($n$)  for the
fixed $n$ model ($\delta = 0$). The symbols represent, as indicated,
analytic results from Onsager's formula and
numerical results. Numerical results are 
for $N \times N = 20 \times 20$. The continous 
curve is the scaling result from Eq.~(\ref{scaling}).
}
\label{fig1}
\end{figure}

In the  heat capacity, features
associated with both
2D and 1D spin fields can be seen - the sharp peak
in $C_{v}$ at $T \approx 0.5$ is
associated with the 2D spin field (and 
the peak position will be henceforth referred to
as $T_{c}^{2D}$) while the 
broad feature in the
heat capacity above $T \approx 0.6$ is associated
with the 1D spin field. The heat capacity eventually
approaches zero at higher temperatures.

For this $\delta=0$ case, we should recover the standard 2D Ising model
results with an effective interaction $J^{(n)}$ 
and this provides an additional check.  Thus,
in Fig.~\ref{fig1}, we plot $T_{c}^{2D}$ vs $n$ based 
on both the analytic calculation (that is, on the Onsager result
for $J^{(n)}$) 
and the numerical simulation. Again, the numerical results
agree with theory.
At $n=0$ we recover the well known 2D 
Ising model transition temperature value,
as expected. As $n$ increases, $T_{c}^{2D}$ decreases indicating 
that  ordering occurs at lower temperatures as
the effective coupling between 2D spins decreases or, 
viewing it in a different way, as
the 1D part of the coupled fields 
becomes more prominent. The relation between $T_{c}^{2D}$ and $n$ can 
also be  obtained (as an alternative to the  $J^{(n)}$ calculation)
via a simple scaling 
argument: the ratio of $n+1$ (the number of
1D links between the nodal spins in the network)
to the 1D Ising correlation length - $\exp(\frac{2}{T})$ (at $T << 1$)-
should remain a constant for all $n$ at the 2D critical
temperature. 
From this scaling argument, we obtain the following 
relation:
\begin{equation}
\label{scaling}
T_{c}^{2D}(n) = \frac{T_{c}^{2D}(n=0)}{1+0.5T_{c}^{2D}(n=0)\log(n+1)}, 
\end{equation}
where $T_{c}^{2D}(n=0)$ is the 2D transition temperature at $n=0$.
This result is plotted as the continuous curve in Fig.~\ref{fig1}.

While comparing the analytic results (which are in the
thermodynamic limit) to the numerical ones, 
for $T_{c}^{2D}$, as in Fig.~\ref{fig1}, 
finite size corrections are inevitably present.
It is shown in Ref.~\onlinecite{fisher} that
for a 2D Ising model, the difference 
in $T_{c}$ between a finite size system ($T_{c}$ for a finite system is
defined to be the temperature at which the heat capacity peaks) and  one in the 
thermodynamic limit is always positive and given by:
\begin{equation}
\label{fisher}
\frac{T_{c}(N)-T_{c}(\infty)}{T_{c}(\infty)} = \frac{a}{N}
\end{equation}
where $a = 0.3603$, $T_{c}(N)$ and $T_{c}(\infty)$ are 
the critical temperatures 
for an $N \times N$ 2D Ising model and in 
the thermodynamic limit respectively. 
For the random coupled field model, 
the result above is modified 
due to the presence of $n$ 1D spins.
The modification to Eq.~(\ref{fisher})
due to $n$ 
can be calculated by rewriting the
equation above in terms of the network model with 
effective interaction
given in Eq.~(\ref{jeff}):
\begin{equation}
\frac{{J^{(n)}(\infty)}/{T}-{J^{(n)}(N)}/{T}}{{J^{(n)}(N)}/{T}} = \frac{a}{N},
\end{equation}
where $J^{(n)}(N)$ denotes $J^{(n)}(T_c(N,n))$, with $T_c(N,n)$ being the 2D
transition temperature for an $N\times N$ 
system with $n$ spins in each link.
It then follows that:
\begin{widetext}
\begin{equation}
\label{finite}
\frac{T_{c}(N,n)-T_c(n)}{T_c(n)} =  
\frac{a [1-\tanh^2({J^{(n)}}/{T_c(n)})] {J^{(n)}} \tanh({1}/{T_c(n)})} 
{Nn \tanh({J^{(n)}}/{T_c(n)}) [1-\tanh^2({1}/{T_c(n)})]} 
\end{equation}
\end{widetext}
where 
$T_c(n) \equiv T_c(\infty,n)$
and $J^{(n)} \equiv J^{(n)}(\infty)$ are the 2D transition temperature and effective
interaction 
in the thermodynamic limit. The $n$ dependence 
in the equation above also 
enters through the effective interaction $J^{(n)}$.
The finite size corrections to $T_{c}^{2D}$ obtained for 
our numerical model agree well with the prediction
in Eq.~(\ref{finite}) above. 
\begin{figure}
\begin{turn}{-90}
\includegraphics[width=0.35\textwidth] {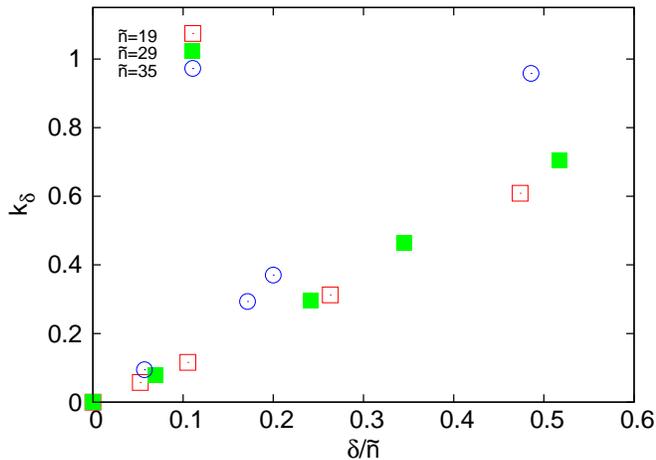}
\end{turn}
\caption{Plot of $k_{\delta}$ (see Eq.~(\ref{parameter}))
vs ${\delta}/{\widetilde{n}}$ for 
three different values of $\widetilde{n}$.}
\label{fig3}
\end{figure}

\subsection{Disorder}
After having validated our procedures
through the fixed $n$ version of our model,
we now 
turn to the random coupled field case. 
We tune the level of disorder in the 
random network by adjusting 
the values $\widetilde{n}$ and $\delta$ 
of the gaussian distribution, Eq.~(\ref{gaussian}).  
Larger values of $\delta$ imply 
a broader gaussian distribution. 
As mentioned above, a useful approach to characterize
the level of  disorder 
in terms of the effective interaction between
2D spins is 
the parameter $k_{\delta}$ as defined in Eq.~(\ref{parameter}).
Since the effective interaction between the 2D spins
depends on $n_{ij}$, randomness in $n_{ij}$ is reflected on $J^{(n)}$ as well. 
In Fig.~\ref{fig3}, we present a plot of 
$k_{\delta}$ vs ${\delta}/{\widetilde{n}}$.
The parameter $k_{\delta}$ which is simply  
the standard deviation of the effective interaction
between 2D spins, 
increases with 
${\delta}/{\widetilde{n}}$ as would be expected, and
it is roughly proportional to it. Note that  
$k_{\delta}$ is also
temperature dependent. This dependence is weak: in Fig.~\ref{fig3} 
we have set the temperature to 
the average annealed value of $T_{c}^{2D}$
for each $\widetilde{n}$. 
\begin{figure}
\begin{turn}{-90}
\includegraphics[width=0.35\textwidth] {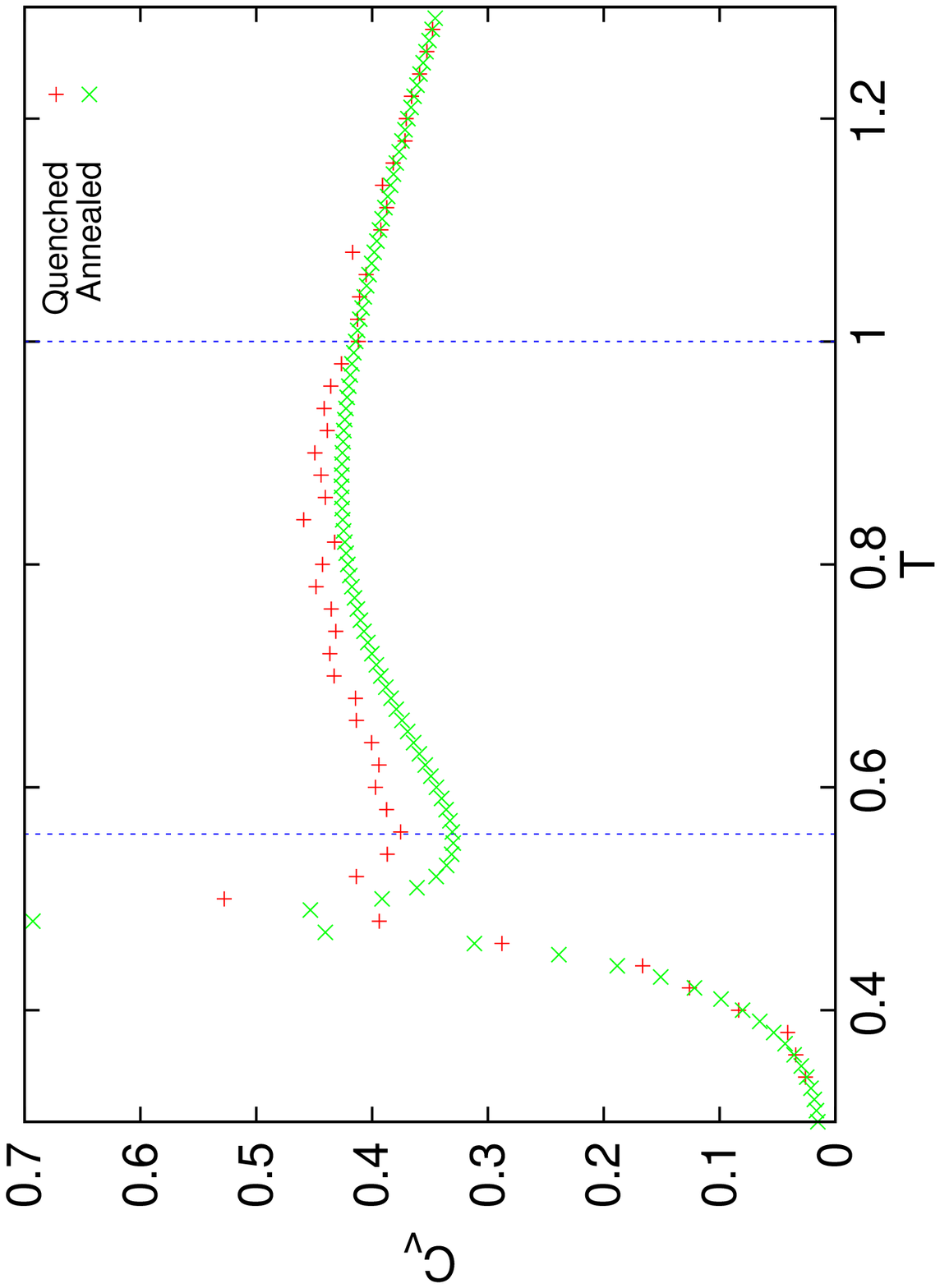}
\end{turn}
\caption{Plot of heat capacity vs temperature for $\widetilde{n}=29$ and $\delta=7$. Dashed lines 
indicate $T_1$ and $T_2$. $T_1$ is the lower temperature limit and $T_2$ the upper
temperature limit in Eq.~(\ref{entr}).}
\label{fig9}

\end{figure}
\begin{figure}
\begin{turn}{-90}
\includegraphics[width=0.35\textwidth] {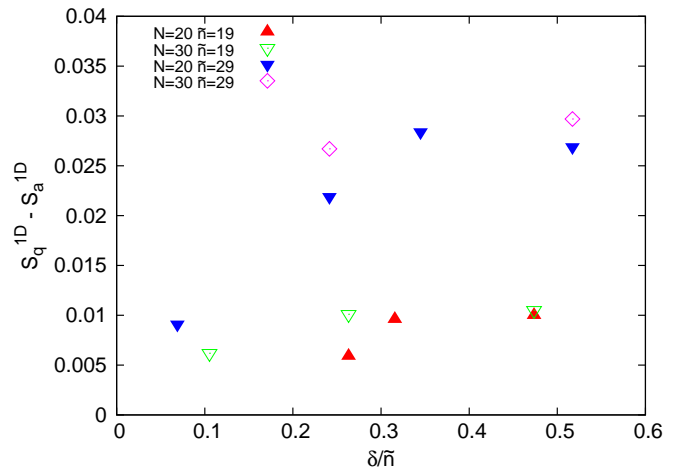}
\end{turn}
\caption{Plot of the difference between quenched and annealed entropy 
associated with 1D fluctuations, as
calculated from Eq.~(\ref{entr}), plotted vs $\delta/\widetilde{n}$.}
\label{fig4}
\end{figure}

\subsubsection{Results for 1D fluctuations}

In quantifying differences between quenched and 
annealed disorder we first look at the 1D field.
In each case, we calculate the  
entropy associated with the 1D fluctuations ($S^{1D}$) from:
\begin{equation}
\label{entr}
S^{1D} = \int\limits_{T_{1}}^{T_{2}} \frac{C_{v}}{T}\mathrm{d}T
\end{equation}
where the temperature limits $T_{1}$ and $T_{2}$ are set,
in order to take into account the 1D contribution to the total heat capacity,
as follows: the lower limit $T_1$ is that of the minimum occurring
between the sharp 2D peak and the broad 1D peak (see
Fig.~\ref{fig9}), while the upper limit $T_2$ 
is taken to be sufficiently high so that there is no longer 
any difference
between the quenched and annealed specific heats (one may therefore
think of $T_2$ as being infinite). 
In Fig.~\ref{fig4}, we plot this difference in the entropies
for the quenched and annealed systems,
associated with  1D fluctuations, for several values of $N$, $\widetilde{n}$
and $\delta$. The weak dependence on $N$ is due to finite size effects
in the numerical calculation for quenched disorder. The 
variation with $\delta/\widetilde{n}$  illustrates the actual dependence of this difference
on the disorder.
We observe that the quenched entropy $S^{1D}_q$ is always greater
than, $S^{1D}_a$, the annealed entropy.
As the level of disorder in the 1D chains in the network increases, 
the entropy difference between the quenched and annealed cases
increases and then saturates at 
${\delta}/{\widetilde{n}}\approx 0.25$. 

\begin{figure}
\begin{turn}{-90}
\includegraphics[width=0.35\textwidth] {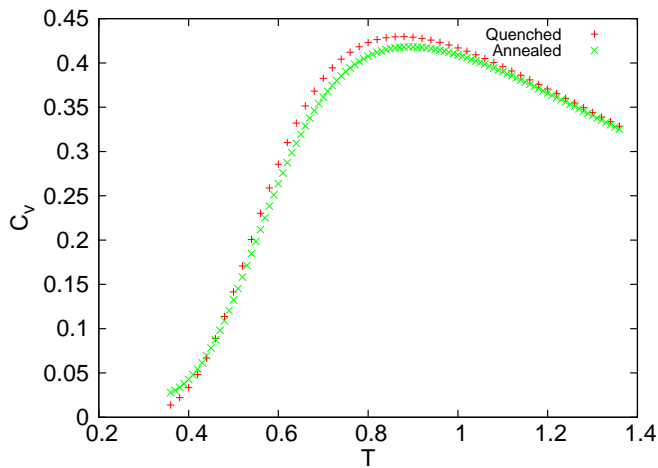}
\end{turn}
\caption{Plot of the contribution of the chains to 
the heat capacity (see text) for quenched and annealed disorder vs. temperature.
The average number of spins in the 1D chains, $\widetilde{n}$, equals 19 
and $\delta=9$.}
\label{fig5}
\end{figure}

The heat capacity due to the 1D chains alone (in the absence of
any 2D spins) can be 
calculated analytically for both quenched and annealed
disorder from Eq.~(\ref{f1}) and the rest of the
discussion in Sec.~\ref{methods}. The temperature dependence of this 1D heat capacity, 
evaluated for both types of disorder, is shown in Fig.~\ref{fig5}. 
For the example plotted there we see that
beginning at $T \approx 0.6$, the 
quenched disorder heat capacity takes on a higher value than the annealed
disorder heat capacity. This mostly accounts for the difference in 
the 1D contribution to the entropy 
of the coupled system, as evaluated  
above from Eq.~(\ref{entr}) and 
plotted in Fig.~\ref{fig4}.
\begin{figure}
\includegraphics[width=0.45\textwidth] {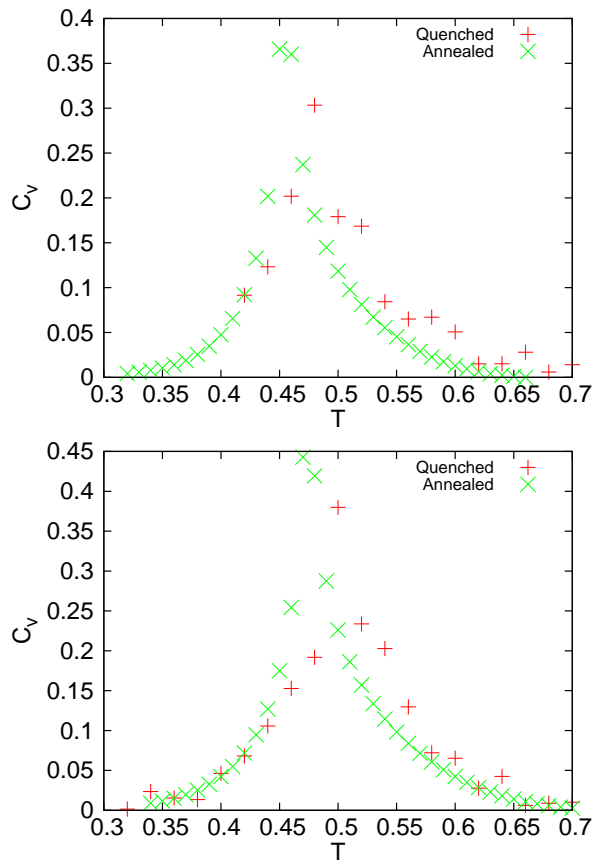}
\caption{The contribution to the
heat capacity from 2D fluctuations (see text),
plotted  vs temperature for both annealed
and quenched disorder. The peak in the heat capacity occurs at $T = T_{c}^{2D}$. 
In the top panel  $\widetilde{n}=35$ and $\delta=2$ 
and in the bottom panel $\widetilde{n}=29$ and $\delta=2$.}
\label{fig7}
\end{figure}
\subsubsection{Results for 2D fluctuations}

We concentrate here on the differences between quenched and 
annealed heat 
capacity due to 2D fluctuations. Since the chain 
contribution to the  heat capacity can be calculated analytically 
for both quenched and annealed disorder (see discussion
in connection with Fig.~\ref{fig5}), we isolate
the 2D contribution to the specific
heat by substracting the chain contribution 
from the total heat capacity. The total $C_v$ is evaluated analytically
in the annealed case and numerically for quenched disorder.
The heat capacity due to 2D fluctuations, as  obtained 
in this manner, is shown in the two panels of 
Fig.~\ref{fig7}, which correspond to
two different sets of values of $\widetilde{n}$ and $\delta$.
We see that the results differ for quenched and
annealed disorder. An important feature of this 
difference  is 
that the 2D transition temperature ($T_{c}^{2D}$) for a frozen 
(quenched) random network is always higher than 
that obtained for the annealed network. 
In the context of our random network of Ising spins, these results 
imply that 
magnetic ordering always takes place at higher temperatures 
for frozen disorder than if the disorder is allowed to anneal. 
In other words, as the time scale associated with the dynamics of the network on which
Ising spins reside changes from being much larger than 
the time scale of spin fluctuations (quenched disorder) to a scenario whereby the two time scales are 
comparable (annealed disorder), the phase transition of the spin system is suppressed.
In terms of the dislocation network problem described in the Introduction, our results
based on a simplified Ising model 
suggest that as the dynamics of the dislocation network become important (i.e. when motion
of dislocation line segments takes place over time scales comparable to those of fluctuations in
the superfluid field), the associated 
phase transition (in this case superfluid ordering) would be suppressed.
Even though superfluid ordering is described by the ferromagnetic XY model, the 
simplified Ising model we have considered 
captures the underlying physical principle: the additional fluctuations present in 
the annealed case will lower the transition temperature. 
\begin{figure}
\begin{turn}{-90}
\includegraphics[width=0.35\textwidth] {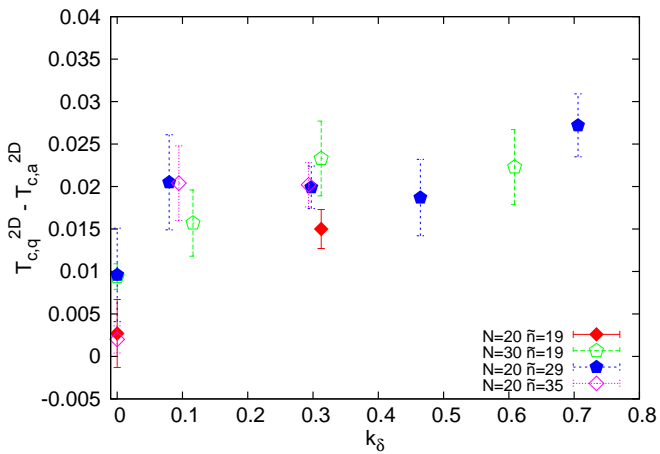}
\end{turn}
\caption{Plot of 
the difference between the
2D transition
temperature (taken to be the temperature at the 2D peak of the specific heat) for 
quenched disorder, $T_{c,q}^{2D}$, and 
the corresponding value for annealed 
disorder, $T_{c,a}^{2D}$. This difference
is plotted vs the parameter $k_{\delta}$ 
(see Eq.~(\ref{parameter})). The error bars
denote numerical uncertainty. The difference between the 2D peak temperatures 
was studied for 
$\widetilde{n} = 19, 29$ and $35$ with a range of values of $\delta$ setting 
the range for $k_{\delta}$.
Numerical results for the quenched case are labelled by the size of the
lattice (N) used in the simulation.} 
\label{fig8}
\end{figure}

In Fig.~\ref{fig8}, we plot the difference in
$T_{c}^{2D}$ between networks with quenched and annealed disorder. 
The error bars arise solely from numerical uncertainties
in the (quenched disorder) numerical results: 
for each point in  Fig.~(\ref{fig8}), the 
quenched 2D transition temperature, $T_{c,q}^{2D}$,
was obtained by averaging over twelve realizations of 
$n_{ij}$ in the 1D chains in the network. 
The error bar associated with each data point
is the standard deviation of the difference in $T_{c}^{2D}$.
It turns
out to be more illuminating to plot
the results for this difference in terms of the
parameter  $k_{\delta}$ (see Eq.~(\ref{parameter})) which characterizes
the width of the distribution of effective couplings, rather than
in terms of the gaussian width, $\delta$, and
average $\widetilde{n}$, of the distribution of $n_{ij}$. 
At $k_\delta\rightarrow 0$ we recover the ordered results: the
difference would be zero in the thermodynamic limit and the
small nonzero results arise from finite size effects in the numerical
calculation: they are described by Eq.~(\ref{finite}). 
The uncertainties due to finite size effects at higher values of 
$k_{\delta}$ remain the same as in the  $\delta \rightarrow 0$ limit.
Earlier studies~\cite{thorpe,falk,thorpe1} on the difference
between annealed and quenched disorder, 
consider the case where the distribution of spin 
interaction strengths is narrow, 
and speculate (without any explicit quenched results) 
that the difference between transition temperatures is small in that limit.
Unlike these earlier studies, our model takes into 
account a broad distribution of effective interaction strengths and we 
obtain explicitly transition temperatures 
for both quenched and annealed models.     
Our results, as seen 
in the region where $k_{\delta} \rightarrow 0$ (corresponding to a narrow distribution of 
interaction strengths) of Fig.~(\ref{fig8}), show that
difference in transition temperature is small in this limit. 
However, we find that 
as the level of 
disorder increases, i.e. for higher values of $k_{\delta}$,
the difference in $T_{c}^{2D}$ between the quenched and annealed 
networks increases rapidly, until 
it saturates at $k_{\delta} \approx 0.1$. 
Beyond  this value
of $k_\delta$ all the points, regardless of the 
varying values of $\widetilde{n}$ and $\delta$ which were
used in the calculation, lie within a narrow
band of values. Thus, it seems indeed that $k_\delta$ 
is sufficient to characterize the phenomena associated
with 2D fluctuations, rather
than $\widetilde{n}$ and $\delta$ separately.
Interpreting this result in the physical context of a network of
dislocation lines, an
increase in $k_{\delta}$ reflects an increase in 
the randomness of a network of dislocation lines due to increasing fluctuation in
dislocation line lengths making up the network. Our results suggest that, as the randomness
in the network increases, the role of the difference in network dynamics (quenched vs annealed)
becomes more important. 

\section{conclusions} 
\label{conclusion}
In this paper, we have studied the role that the type of
disorder - quenched or annealed - plays in the thermodynamic
behavior of  
an Ising model defined on a random network. This
network consists of four-fold coordinated Ising spins connected
by spin chains. The
strength of the disorder can be tuned by varying the
average value of the chain length and its standard
deviation. We have emphasized both
the transition temperature and the specific heat in
the region dominated by one dimensional fluctuations.
We have shown that the  
transition temperature for our Ising model on a random network
in which the disorder is quenched (frozen) is always higher than the
transition temperature for annealed disorder with the same distribution.
The magnitude of the difference between the two transition temperatures
is quantified by our study.
We also show that the entropy associated with the one dimensional
fluctuations is larger for the quenched case.

Our study quantifies the difference between the properties
of quenched and annealed versions of disordered systems.
The quenched assumption applies 
when  the time scale over which the disorder changes is much longer
than that for the spin fluctuations.
In our model
the strength of the effective interaction between neighboring sites
has a very broad distribution.
Its general interest is that it relates to a variety of experimentally 
studied systems
in which the strength of the effective interaction between neighboring spins
has a very broad distribution.
An example is dilute 
magnetic semiconductors.
Our results indicate that the transition 
temperature and other thermodynamic properties of dilute 
magnetic semiconductors might be approximated
from an analytic calculation for the 
annealed model with
the same distribution
of interaction strengths.  
Our model may be relevant also to the renewed  interest on 
dislocation networks in solid $^4$He. It presents a 
simplified version of
how the dynamics of a dislocation network may
influence a superfluid field in its vicinity.
Our results indicate that in the annealed scenario, 
when fluctuations of dislocation line segments 
within a network become important i.e. when the time scale
for dislocation line fluctuations becomes comparable to or smaller than 
the time scale associated with fluctuations of the superfluid field, 
the associated phase transition is suppressed.  
On the other hand, superfluid ordering would 
be enhanced in the vicinity of a dislocation 
if the dislocation 
network can be considered to be frozen. While our results have been obtained for a simplified
Ising version of the superfluid transition, we expect that the general conclusion about the transition
being suppressed by fluctuations in the dislocation network will
remain valid when the proper symmetry of the superfluid order parameter (XY model) is taken
into account. Quantum effects, 
considered in Ref.~\onlinecite{balibar}, 
but not taken into account in our study, are
expected to enhance the suppression of the superfluid transition by the motion of dislocation
lines. 

\acknowledgments This research was supported in part by IUSSTF grant 94-2010.
A.M.K. would like to thank Sumanta Bandyopadhyay and Cole Grasse
for their help.

\end{document}